\documentstyle[aps,prd,twocolumn,epsf]{revtex} 

\draft

\begin{document}

\title{Imprint of Sterile Neutrinos in the Cosmic Microwave Background
Radiation}

\author{Steen Hannestad}

\address{Theoretical Astrophysics Center, 
Institute of Physics and Astronomy,
University of Aarhus,
DK-8000 \AA rhus C, Denmark}

\author{Georg Raffelt}

\address{Max-Planck-Institut f\"ur Physik 
(Werner-Heisenberg-Institut),
F\"ohringer Ring 6, 80805 M\"unchen, Germany}

\date{Revised Version: September 30, 1998}

\maketitle

\begin{abstract}
  The existence of low-mass sterile neutrinos is suggested by the
  current status of solar and atmospheric neutrinos together with the
  LSND experiment. In typical four-flavor scenarios, neutrinos would
  contribute to a cosmic hot dark matter component {\em and\/} to an
  increased radiation content at the epoch of matter-radiation
  equality. These effects leave their imprint in sky maps of the
  cosmic microwave background radiation (CMBR) and may thus be
  detectable with the precision measurements of the upcoming MAP and
  PLANCK missions.
\end{abstract}



\section{Introduction}

Neutrino oscillations are currently indicated by the
solar~\cite{solar} and atmospheric~\cite{atmospheric} neutrino
anomalies and by the LSND experiment~\cite{LSND}.  Taken together,
these three bits of evidence are too much of a good thing in that they
are incompatible with a three-flavor mixing scheme among $\nu_e$,
$\nu_\mu$ and $\nu_\tau$. Apart from the obvious possibility that some
of these preliminary indications may be unrelated to neutrino
oscillations, one intriguing speculation is that there is a fourth
low-mass neutrino, $\nu_s$, which mixes with the standard
flavors~\cite{fourflavor}.  It would have to be sterile with regard to
the electroweak interactions and thus is undetectable in any direct
search experiment.

The mixing of $\nu_s$ with standard flavors allows for its thermal
production in the early universe, and even though it will typically
not attain full equilibrium there will be a cosmic background of
sterile neutrinos. The standard Big Bang nucleosynthesis (BBN)
constraint on the cosmic radiation density thus provides nontrivial
limits on the masses and mixing angles of a four-neutrino scenario
consisting of $\nu_e$, $\nu_\mu$, $\nu_\tau$ and $\nu_s$
\cite{BBN1,Viking,BBN2,sarkar}.  Likewise, if neutrinos are Dirac particles
and thus have right-handed components and if they have anomalous
magnetic dipole moments, a cosmic abundance of the sterile states can
be produced by magnetically induced spin precessions and by
electromagnetic spin-flip scatterings~\cite{dipole}.

However, the most spectacular cosmological consequence of sterile
neutrinos is their impact on the large-scale structure of the
universe, and notably on the temperature variations of the cosmic
microwave background radiation (CMBR).  The anticipated sky maps of
the future MAP and PLANCK~\cite{MAP+PLANCK} satellite missions have
already received advance praise as the ``Cosmic Rosetta
Stone''~\cite{BTW} because of the wealth of cosmological precision
information they are expected to
reveal~\cite{CMBreview,Jungman,Bond,HETW}.  In the previous discourse
on sterile neutrinos it has been curiously overlooked that a
successful deciphering of the CMBR hieroglyphs could well make or
break the hypothesis of this elusive particle's existence. Even if its
signature in real CMBR sky maps may not be unambiguously visible, the
hypothesis of sterile neutrinos introduces two additional degrees of
freedom into the game of cosmological parameter estimation, viz.\ a
hot dark matter component and additional radiation in the form of
neutrinos.


\section{Sensitivity to Radiation Content}

CMBR sky maps are characterized by their fluctuation spectrum
$C_\ell=\langle a^{}_{\ell m} a^*_{\ell m}\rangle$ where $a_{\ell m}$
are the coefficients of a spherical-harmonic expansion. Fig.~1 (solid
line) shows $C_\ell$ for standard cold dark matter (SCDM) with $h=0.5$
for the Hubble constant in units of $100~{\rm km~s^{-1}~Mpc^{-1}}$,
$\Omega_M=1$ and $\Omega_B=0.05$ for the matter and baryon content, a
Harrison-Zeldovich spectrum of initial density fluctuations, ignoring
reionization, and taking $N_{\rm eff}=3$ for the effective number of
thermal neutrino degrees of freedom.

Sterile neutrinos increase the radiation content and thus modify this
pattern in a characteristic way illustrated by the dotted line in
Fig.~1 which corresponds to $N_{\rm eff}=4$.  While this shift appears
small, the lower panel of Fig.~1 shows that for $\ell\agt 200$ it is
large on the scale of the expected measurement precision. It is
fundamentally limited by the ``cosmic variance'' $\Delta
C_\ell/C_\ell=\sqrt{2/(2\ell+1)}$, i.e.\ by the fact that at our given
location in the universe we can measure only $2\ell+1$ numbers
$a_{\ell m}$ to obtain the expectation value $\langle a^{}_{\ell m}
a^*_{\ell m}\rangle$.  The actual sensitivity will be worse, but the
cosmic variance gives us an optimistic idea of what one may hope to
achieve.

The true sensitivity to $\Delta N_{\rm eff}$ is further limited by our
lack of knowledge of several other cosmological parameters.  Even then
it is safe to assume that we are sensitive to $|\Delta N_{\rm
eff}|\alt0.3$, and much better with prior knowledge of other
parameters~\cite{Jungman}. Thus it is clear that the CMBR is a 
more powerful tool to measure $N_{\rm eff}$ than the standard BBN
argument which informs us that $|\Delta N_{\rm eff}|\alt 1$, where the
exact limit adopted by various authors depends on their attitude
towards the systematic uncertainties of the primordial light-element
abundances~\cite{BBN3}.

The most optimistic assessment of the $\Delta N_{\rm eff}$ sensitivity
that may be achieved with future CMBR experiments was recently put
forth in Ref.~\cite{Lopez}.  It was claimed that without polarization
measurements and without priors of other cosmological parameters one
could see $\Delta N_{\rm eff}\alt0.4$ if the experiment measures on
angular scales up to $\ell_{\rm max}=1000$ (roughly corresponding to
MAP), and $\Delta N_{\rm eff}\alt0.1$ for $\ell_{\rm max}=2000$
(roughly PLANCK).  With polarization measurements one improves to
$\Delta N_{\rm eff}\alt0.1$ (MAP) and 0.04 (PLANCK), while including
priors achieves 0.02 and 0.008, respectively.  With both polarization
measurements and priors available one could reach $\Delta N_{\rm
eff}\alt0.008$ (MAP) and 0.002 (PLANCK), taking us truly into the
realm of precision cosmology!

\begin{figure}[ht]
\vskip4pt
\hbox to\hsize{\hss
\epsfxsize=8.45truecm\epsfbox{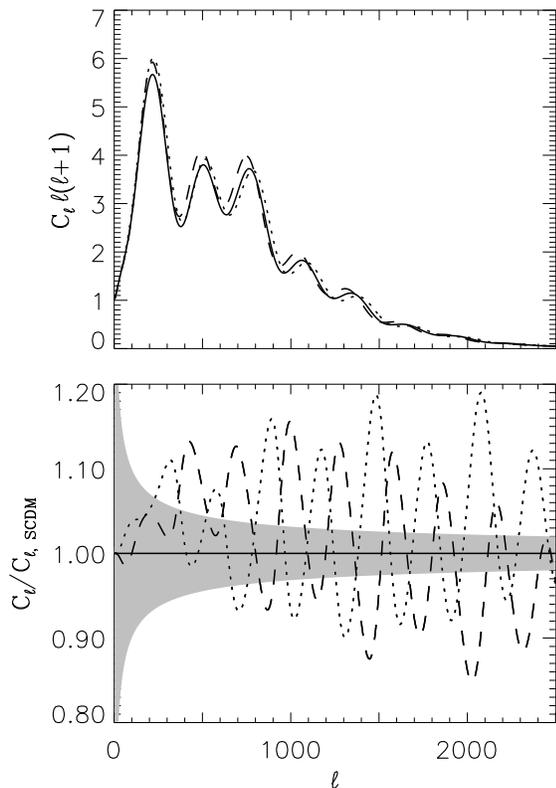}\hss}
\bigskip
\vspace*{0.5cm}
\caption{{\it Top:} CMBR fluctuation spectrum for SCDM with $h=0.5$,
  $\Omega_M=1$, $\Omega_B=0.05$, and $N_{\rm eff}=3$ (solid line). The
  dotted line is for $N_{\rm eff}=4$, and the dashed line when two of
  these four neutrinos have equal masses corresponding together to
  $\Omega_{\rm HDM}=0.2$ ($\Omega_{\rm CDM}=0.75$).  {\it
  Bottom:}~Relative difference of these nonstandard models to SCDM.
  The shaded band represents the cosmic variance.  (Spectra calculated
  with the CMBFAST~\protect\cite{CMBFAST} package.)}
\label{fig1}
\end{figure}

There are several reasons why these assessments are probably overly
optimistic. First, the interpretation of the CMBR signal may be
significantly affected by foreground emissions. The treatment in
Ref.~\cite{Lopez} assumes that the primary error in the data will
be due to cosmic variance and neglects possible foreground contamination.
This is a problem
which can only be treated properly once the new data become available
since the nature and magnitude of possible foregrounds are not well
known at present (for a discussion see Ref.~\cite{tegmark}).  Second,
the explored cosmological parameter space is limited. There are
``degeneracies'' between the effect of varying several of the dozen or
so standard cosmological parameters which determine the CMBR sky maps.
These degeneracies can be broken by other observations,
for example the anticipated galaxy correlation functions from the
Sloan Digital Sky Survey (SDSS)~\cite{sloan}.  In the most recent
analysis~\cite{HETW} it was claimed that PLANCK-level CMBR
observations with polarization information together with SDSS will
achieve only a precision of $\Delta N_{\rm eff}\alt0.2$ at the
1$\sigma$ level. According to this assessment it will be a struggle
to beat the BBN precision of the $N_{\rm eff}$ determination. 

In our following discussion we will take the attitude that a $\Delta
N_{\rm eff}$ of a few 0.1 will be detectable, and that a value as
small as 0.01 is not ignorable for the cosmological parameter
estimation, even if it may not be identifyable from the CMBR sky maps.


\section{Massless Neutrinos}

As a simple generic case we begin with a four-flavor scenario where
the masses are so small that all neutrinos are ultra-relativistic at
the epoch of matter-radiation equality ($T_{\rm eq}=5.5~{\rm
eV}~\Omega_M h^2$), i.e.\ $m_\nu\ll 1~{\rm eV}$. This implies that the
only cosmological effect of $\nu_s$ is its contribution to $N_{\rm
eff}$.

Calculations of $N_{\rm eff}$ from primordial
$\nu_e$-$\nu_s$-oscillations as a function of the assumed masses and
mixing angles have been performed by many
authors~\cite{BBN1,Viking,BBN2}; we follow the simple method of
Ref.~\cite{Viking}. The neutrino ensemble is characterized by a single
flavor-polarization vector, i.e.\ the entire ensemble is treated as
having the average momentum $\langle p \rangle = 3.15\,T$. As long as
there are no resonant oscillations this is sufficiently accurate since
neutrinos are kept in kinetic equilibrium until long after they
decouple from chemical equilibrium.  In the case of resonant
transitions the situation is complicated by the fact that different
momentum modes pass through the resonance at different temperatures.

Figure~2 shows our results for the equivalent number of extra light
neutrinos, $\Delta N_{\rm eff}$, as a function of the oscillation
parameters $\sin^22\theta$ and $\delta m^2$ where we have taken
$m_{\nu_s}>m_{\nu_e}$.  Also shown are the 95\% C.L.\ regions for the
sterile-neutrino MSW solutions of the solar neutrino
problem~\cite{bahcall} and the $3\sigma$ favored solution of the
atmospheric neutrino anomaly from
$\nu_\mu$-$\nu_s$-oscillations~\cite{FVY98}.

The solar small-angle MSW solution would correspond to a $\Delta
N_{\rm eff}$ at the $10^{-3}$ level, which is undetectable even under
the most optimistic assumptions. Likewise, the vacuum solution at
$\delta m^2\approx 10^{-10}~{\rm eV}^2$ has no impact whatsoever on
the~CMBR.

The large-angle MSW solution would correspond to $\Delta N_{\rm
eff}\approx 0.1$, perhaps too small to be clearly visible in the CMBR
sky maps. However, it could not be ignored when determinining the
other cosmological parameters.

\begin{figure}[b]
\hbox to\hsize{\hss
\epsfxsize=7.51truecm\epsfbox{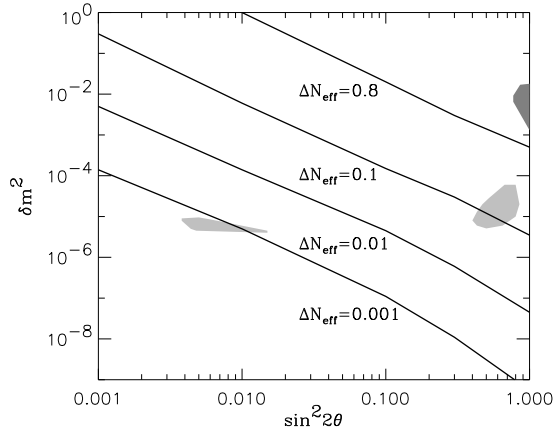}\hss}
\bigskip
\caption{$\Delta N_{\rm eff}$ caused by primordial
$\nu_e$-$\nu_s$-oscillations with $m_{\nu_s}>m_{\nu_e}$.  Also shown
are the 95\% C.L.\ allowed regions for solar small- and large-angle
MSW $\nu_e$-$\nu_s$-oscillations \protect\cite{bahcall} (light shade)
and the 3$\sigma$ allowed region for atmospheric
$\nu_\mu$-$\nu_s$-oscillations \protect\cite{FVY98} (dark shade).}
\label{fig2}
\end{figure}

The atmospheric neutrino anomaly can be explained by
$\nu_\mu$-$\nu_s$-oscillations with nearly maximum mixing and $\delta
m^2=10^{-3}$--$10^{-2}~{\rm eV}^2$ as indicated by the dark-shaded
region in Fig.~2. While the contours where calculated for
$\nu_e$-$\nu_s$-oscillations, they roughly also apply to the present
case if $m_{\nu_s}>m_{\nu_\mu}$. We have checked that independently of
the sign of $\delta m^2$ the sterile neutrinos reach almost perfect
thermal equilibrium so that a $\nu_\mu$-$\nu_s$-solution of the
atmospheric neutrino anomaly should stick out clearly from the CMBR
data.  This can be seen in Fig.~2 where the atmospheric solution
yields a $\Delta N_{\rm eff} > 0.8$, even for non-resonant
oscillations.

It deserves mention that a sterile species can be thermally excited by
other mechanisms than a mass term.  For instance, if the neutrino had
a Dirac magnetic dipole moment, the right-handed components can be
brought into thermal equilibrium by spin-flip interactions with the
electromagnetic plasma~\cite{dipole}.  Using the CMBR one should
therefore be able to constrain the neutrino Dirac dipole moment
somewhat tighter than with BBN. Likewise, extra radiation can be
produced by exotic neutrino decays of the sort $\nu\to\nu'\phi$ with
$\phi$ a new massless boson such as the Majoron. One of us has already
explored the imprint of such scenarios on CMBR sky
maps~\cite{Hannestad}.


\section{Hot plus Cold Dark Matter (HCDM)}

The LSND experiment indicates a mass difference between $\nu_e$ and
$\nu_\mu$ of anywhere between about 0.4 and 3~eV~\cite{LSND}. Taking
this result as well as the solar and atmospheric anomalies as serious
indications for neutrino oscillations leads us naturally to a
four-flavor scenario with two neutrino pairs, each consisting of two
nearly mass-degenerate states, and with an eV-range mass separation
between the pairs~\cite{fourflavor}.  This would imply that neutrinos
play a cosmological role as a hot dark matter (HDM) component and as
such correct the problem of overproducing small-scale structure which
bedevils SCDM models~\cite{Primack}. The small-scale power spectrum of
the cosmic matter-density fluctuations will be measured with
unprecedented precision by the Sloan Digital Sky Survey \cite{sloan}.
It was recently shown that these measurements may well be sensitive
down to the lower end of LSND-inspired neutrino masses~\cite{Hu}.

In addition, there would be an imprint in the CMBR fluctuation
spectrum~\cite{Ma}. Neutrinos with eV masses are still relativistic at
the epoch of matter-radiation equality so that the HDM component in a
HCDM scenario initially counts toward the cosmic radiation density,
and only later to the matter density.  Essentially, by giving mass to
the neutrinos we have removed matter from the CDM component when
holding $\Omega_M=1$ fixed so that adding neutrino masses mimics extra
radiation at the epoch of matter-radiation equality in a standard flat
CDM model.  This enhances the first Doppler peak via the early
integrated Sachs-Wolfe effect in analogy to extra
radiation~\cite{tegmark}.  Of course, beyond the first peak the
modification is more intricate, but the main physical effect at large
angular scales can be understood in this way.

In an optimistic interpretation of what PLANCK may achieve, the
sensitivity to a HDM component may be as good as $\delta \Omega_{\rm
HDM}\alt 0.02$~\cite{Bond}.  In a 2$\nu$CDM picture (two
mass-degenerate neutrinos as HDM component) we have
$\Omega_{2\nu}h^2=2m_\nu/93~{\rm eV}$, implying an optimistic PLANCK
sensitivity to a neutrino mass as low as $m_\nu\alt 0.25~{\rm eV}$ if
$h=0.5$.

HCDM scenarios remedy the SCDM problem of overproducing small-scale
structure, but there are other possible solutions to this problem.
Therefore, the primary motivation for a HDM component of eV-mass
neutrinos arises from the LSND measurements which in turn suggest a
sterile neutrino if the solar and atmospheric indications are taken
seriously as well. (In order to avoid sterile neutrinos, many authors
would rather discard the LSND results than any of the other two hints
for oscillations; the conflict with the KARMEN limits~\cite{KARMEN} is
getting difficult to ignore.)  As a consequence, four-flavor neutrino
mass schemes and HCDM scenarios are closely intertwined hypotheses.

For example, if the atmospheric neutrino anomaly is due to
$\nu_\mu$-$\nu_s$-oscillations, we will have approximately $N_{\rm
eff}=4$, and two of these states will have an eV-range mass. The CMBR
imprint of this scenario is illustrated with the dashed curve in
Fig.~1 where we have chosen $\Omega_\nu=0.2$.  With
$\Omega_{2\nu}h^2=2m_\nu/93~{\rm eV}$ and taking $h=0.5$ this implies
$m_\nu\approx2.4~{\rm eV}$, well within the range suggested by~LSND.
This value for $\Omega_\nu$ gives the best fit to observations of the
large scale structure, as noted by several authors~\cite{Primack}.

The region around the first acoustic peak is seen to be enhanced
substantially compared with the massless $N_{\rm eff}=4$ scenario.  As
explained earlier, giving mass to the neutrinos mimics the effect of
extra radiation, at least around the first acoustic peak, so that
in an $N_{\rm eff}=4$ scenario with massive neutrinos the separate
effects add to a larger compound imprint. 

Other four-flavor scenarios have a less dramatic impact, notably if
the sterile state solves the solar neutrino problem with a small
mixing angle or a very small mass difference to $\nu_e$. Still, in any
of the data-inspired four-flavor schemes one cannot avoid worrying
about both, a HDM component and extra radiation.

For any given mass and mixing scheme one can work out $N_{\rm eff}$
and the HDM component. However, this can be a complicated task when
resonant effects become important which, in turn, depend on the
unknown primordial lepton-number asymmetry.  It has been shown that
resonant oscillations can generate a significant $\nu_e$-$\bar\nu_e$
asymmetry which affects the primordial helium production through
modified $\beta$ reaction rates~\cite{BBN2}. Therefore, in four-flavor
scenarios, BBN is not always a faithful probe for the radiation
content which we express in terms of $N_{\rm eff}$. Put another way,
the BBN-quantity $N_{\rm eff}$ is an indirect measure of the helium
yield, while our $N_{\rm eff}$ is a measure of the radiation content
at the epoch of matter-radiation equality. The two notions can be
vastly different and are separately important. 
The main point here is that BBN is sensitive to the flavour of neutrinos
whereas the CMBR measures only energy density.


\section{Conclusion}

Low-mass sterile neutrinos are a generic
possibility, and indeed required if all current empirical indications
for neutrino oscillations are correct.  This would imply a
cosmological hot dark matter component in the form of massive
neutrinos, and nonstandard contributions to the radiation density at
the epoch of matter-radiation equality.  In contrast with previous
discussions, both effects would simultaneously occur and would leave
their imprint in the large-scale matter distribution as well as in the
CMBR temperature sky~maps.

In a four-flavor scenario, the neutrino mass- and mixing scheme can be
rather complicated, allowing for involved oscillation phenomena in the
early universe because of the possibility of resonant effects. It is
thus premature to attempt a complete discussion of all possible
cases. However, if one takes the current empirical situation with
regard to neutrino parameters seriously at all, then nonstandard
neutrino properties will have a large impact on the cosmological
observables to be extracted from precision CMBR experiments and galaxy
surveys.  In some scenarios, the sterile-neutrino imprint will stick
out very clearly, in others it may not be possible to disentangle it
from other effects. The most difficult-to-detect scenario is where
atmospheric neutrinos oscillate from $\nu_\mu$ to $\nu_\tau$ and solar
neutrinos from $\nu_e$ to $\nu_s$ with the small mixing angle MSW
solution or the vacuum solution. 

Even if the signature of sterile neutrinos cannot be unambiguously
seen in the CMBR sky maps and galaxy surveys, they still affect the
interpretation of these cosmological precision observables.
Therefore, the current experimental effort to pin down the neutrino
mass spectrum and mixing angles is inseparably interwoven with a
precision interpretation of the forthcoming CMBR~sky~maps.


\acknowledgements

We thank T.~Weiler for informative
discussions of four-flavor neutrino scenarios.  This work was
supported by the Theoretical Astrophysics Center under the Danish
National Research Foundation and by the Deutsche
For\-schungs\-ge\-mein\-schaft under grant No.\ SFB~375.


\end{document}